# A pixel-based approach to massive lesion detection in X-ray mammography

I.Gori[a,*], A. Retico[b]

[a]*Bracco Imaging S.p.A., Via E. Folli 50, 20134 Milano, Italy*
[b]*Istituto Nazionale di Fisica Nucleare, Sezione di Pisa, Largo B. Pontecorvo 3, 56127 Pisa, Italy*

*Keywords:* Computer-aided detection (CAD); Mammographic screening; Neural Networks

## 1. Introduction

A system for the automated detection of massive lesions in mammograms is presented. The approach we adopted is a pixel-based and multi-level one [1, 2]. Each pixel in a mammogram is flagged with the appropriate class membership, e.g. massive lesions or normal breast tissue.

## 2. Methods

The pixel-based detection scheme we propose is referred as Pixel-Matrix Theory (PMT). It is based on the following main steps: a) a vector of *features* extracted from the matrix of the grey-level values of a pixel and its neighbourhood (minimum size 3x3) is assigned to each pixel of the mammogram; b) the vector of *features* is analysed by a neural classifier, which assigns to the pixel a measure of suspiciousness; c) the pixels are grouped in connected regions and a method for reducing the amount of false-positive findings is implemented. Within the PMT framework the *features* characterizing the pixel are the singular values of the intensity matrix, the eigenvalues and the eigenvectors of the gradient matrix, the eigenvalues of the Hessian matrix. A multi-level analysis is performed: the mammogram is analyzed at several resolution scales and the *features* are extracted from matrices of different sizes.

## 3. Results and Conclusion

A set of 70 mammograms (45 containing massive lesions and 35 constituted only by normal tissue) were used for a preliminary test of the method. The best performances achieved in classifying the pixels are 87.9% for the sensitivity and 88.0% for the specificity. An algorithm for reducing the false positive findings generated by the system is currently being tested. The preliminary results are satisfactory and an enhancement of the performances is expected.

## References

[1] W.P. Kegelmeyer, J.M. Pruneda, P.D. Bourland et al., Computer-aided mammographic screening for speculated lesions, Radiology 191: 331-337 (1994).
[2] G. te Brake, N. Karssemeijer, Single and Multiscale Detection of Masses in Digital Mammograms, IEEE T Med Imaging 18(7): 628-639 (1999).

---

[*] Corresponding author. Tel.: +39-0502214397; fax: +39-0502214460.
*E-mail address*: Ilaria.Gori@bracco.com